\documentstyle{article}
\input{psfig}
\newcommand{\beq}{\begin{equation}}
\newcommand{\eeq}{\end{equation}}
\newcommand{\beqa}{\begin{eqnarray}}
\newcommand{\eeqa}{\end{eqnarray}}
\newcommand{\la}{\langle}
\newcommand{\ra}{\rangle}
\newcommand{\bv}{\hat b}
\newcommand{\dtwo}{{\delta^2}}
\newcommand{\ELJ}{E_{\mbox{\tiny LJ}}}
\newcommand{\Tr}{\mbox{Tr}\;}

\newcommand{\ie}{{\it i.e.\ }}
\newcommand{\lsim}{\:\raise -4pt\hbox{$\stackrel{\textstyle <} {\sim}$}\:}

\begin{document}
\begin{titlepage}
\begin{flushright}

LU-TP 97-04 \\
\today \\

\end{flushright}
\vspace{0.0in}
\LARGE
\begin{center}
{\bf On the Kinetic Behavior and Folding Properties \\
of an Off-Lattice Heteropolymer Model}\\
\vspace{0.5in}
\large
Ola Sommelius\footnote{ola@thep.lu.se} \\
\vspace{0.2in}
Complex Systems Group, Department of Theoretical Physics\\
University of Lund, S\"olvegatan 14A, S-223 62 Lund, Sweden\\
\vspace{0.5in}

Submitted to {\em Europhysics Letters}

\end{center}
\vspace{0.2in}

\large
{\bf Abstract:} \\
The kinetic behavior of a three-dimensional off-lattice heteropolymer model is
studied in terms of the time dependence of the average mean-square displacement 
between configurations. It is found that at short time-scales similar behavior
is obtained even for sequences with very different thermodynamic properties. 
Furthermore, the degree of cooperativity in the folding process is examined
by studying  the residual number of degrees of freedom, obtained from an 
eigenvalue analysis of the correlation matrix, contributing to 
the structural fluctuations. In the compact state, a gradual decrease in
this effective number of degrees of freedom take place as the temperature
is lowered. This can be interpreted as an increasing asymmetry of the energy 
landscape.
\end{titlepage}

%\large
\section{Introduction}
In order to understand the characteristics of the energy landscape and to 
quantify the requirements needed for a protein to have a thermodynamical 
stable yet kinetically accessible native state, the introduction of simple 
models \cite{Karplus:95} is necessary. In these models the small length and 
time scales are effectively averaged out and the thermodynamics of the 
resulting coarse grained chain is described by an effective potential. The 
basic assumption for this is that a low-resolution description still captures 
features essential for describing folding properties.

Models which have been much studied are the lattice models where the protein is 
represented as chain of beads on a cubic lattice (see references in 
\cite{Karplus:95}). This approach has indeed proven to be very useful and one
is able to describe several non-trivial aspects in the folding process. 
However it is important to study alternative off-lattice models both in order 
to understand the limitations of the lattice models and in their own right.
In this work a 3D off-lattice model suggested in \cite{Irback:96} is used. 
The model contains two types of 
amino acids, hydrophobic and hydrophilic, and the formation of a hydrophobic 
core is induced by a sequence dependent Lennard-Jones potential.  
In \cite{Irback:96} the studies were focused on the thermodynamical behavior of 
this model.

For a sequence to be a good folder it must satisfy both 
thermodynamic and kinetic requirements. That is the ground state must be 
stable against thermodynamical fluctuations, which happens at low 
temperatures, and yet be kinetically accessible, which happens at higher
temperatures. Therefore only sequences for which the thermodynamic stability
persists at high enough temperatures can be classified as good folders. 
To understand the behavior of the structural fluctuations in this context
the overall shape diffusion at short times is studied as well as the collective
behavior of the conformational correlations in the chain.

This letter is organized as follows. Section two defines the model and the 
observables. Section three and four contain the results and a brief summary 
respectively.
\section{The model}
The model contains two kinds of residues, $A$ and $B$, which behave
as hydrophobic  ($\sigma_i=$'$A$') and hydrophilic ($\sigma_i=$'$B$') 
residues, respectively. These monomers are joined by rigid bonds $\bv_i$
to form a linear heteropolymer chain living in three dimensions. The shape 
of the chain is thus specified by the $N-1$ bond vectors $\bv_i$.
The energy function is given by
\beq
E(\bv; \sigma) =  -\kappa_1\sum_{i=1}^{N-2}\bv_i\cdot\bv_{i+1}  
- \kappa_2\sum_{i=1}^{N-3}\bv_i\cdot\bv_{i+2} + 
\sum_{i=1}^{N-2}\sum_{j=i+2}^N \ELJ(r_{ij};\sigma_i,\sigma_j)
\label{energy}
\eeq   
where $r_{ij}=r_{ij}(\bv_{i},\ldots,\bv_{j-1})$ denotes the distance between 
sites  $i$ and $j$ of the chain, and $\sigma_1,\ldots,\sigma_N$ is a binary 
string that specifies the primary sequence. The {\it species-dependent} global 
interactions are given by the Lennard-Jones  potential,    
\beq
\ELJ(r_{ij};\sigma_i,\sigma_j)=
4\epsilon(\sigma_i,\sigma_j)\Big( \frac{1}{r_{ij}^{12}}-
\frac{1}{r_{ij}^{6}}\Big)\ .
\label{lj}
\eeq
The depth of the minimum of this potential, $\epsilon(\sigma_i,\sigma_j)$, 
is chosen to favor the formation of a core of $A$ residues, \ie 
$\epsilon(A,A) = 1$ and $\epsilon(A,B) = \epsilon(B,B) = 1/2 $.
The two parameters of the energy function, $\kappa_1$ and $\kappa_2$, 
determine the strength of {\it species-independent} local interactions. 
The thermodynamic behavior of this model has been studied in 
Ref.~\cite{Irback:96}. It was found that the values 
$(\kappa_1,\kappa_2) = (-1, 0.5)$ give rise to local correlations 
qualitatively similar to those found  in functional proteins.
This choice favors anti-parallel nearest neighbour bonds and parallel next 
nearest neighbour bonds and will be used throughout this work.
Furthermore it was found that, as the temperature is lowered, a gradual 
compaction occur. In the compact state this is then followed by a 
sequence dependent folding transition.
For this model only a small fraction of random sequences  have good 
folding properties. In 2D \cite{Irback:97}, where a more extensive investigation
was done, only around 10  percent satisfied the folding criteria.
This model to some extent resembles the IMP model \cite{Iori:91, Marinari:92a} 
which is a Gaussian chain augmented with a Lennard-Jones potential with an 
additional quenched disorder term $\sqrt{\epsilon}\,\eta_{ij}/r_{ij}^6$ 
representing a species-dependent interaction. The $\eta_{ij}$'s are stochastic
variables having zero mean and unit variance, while $\epsilon$ is a measure
of the strength of the quenched disorder.
%
%\subsection{Observables}

In order to study the kinetics on a rugged energy landscape some kind of 
distance measure between conformations has to be defined. A natural 
choice \cite{Iori:91} is the mean-square displacement between two 
configurations 
$a$ and $b$, $\delta^2_{ab}$:
\beq
\delta^2_{ab} = \min \frac{1}{N} \sum_{i=1}^{N}|{\bf x}_i^a -  {\bf x}_i^b |^2
\label{d2}
\eeq 
where ${\bf x}_i^{a(b)}$ denotes the position of monomer $i$ in system 
$a$($b$). The minimization is to be performed over translations, rotations and 
reflections. 

With  $\delta_0^2$ denoting the distance to the ground state, and
 $P(\delta_0^2)$ the corresponding probability distribution,  we can define the 
probability for the system to be found in the vicinity of the ground state as:
\beq
p_0 = \int_0^{0.04} P(\delta_0^2)d\delta_0^2
\eeq
The folding temperature $T_f$ is then defined as the temperature where 
$p_0 = 1/2$.
\section{Results}
We use six sequences, of length $N=20$ (Table \ref{tab:1}),  chosen in 
\cite{Irback:96} to represent a variety of behavior, and examine their kinetic
properties for short (intermediate) times. That is, the times should be large 
enough so as not depend on details of the  MC-method (here the normal Metropolis
algorithm) but smaller than the time scales necessary to equilibrate the 
systems. 
\begin{table}[b]
\begin{center}
\begin{tabular}{|c|c|c|} \hline
no.      & sequence                         & $T_f$ \\
\hline
1        & $BAAA\ AAAB\ AAAA\ BAAB\ AABB$   & $<0.15$ \\
2        & $BAAB\ AAAA\ BABA\ ABAA\ AAAB$   & $<0.15$ \\
3        & $AAAA\ BBAA\ AABA\ ABAA\ ABBA$   & 0.23    \\
4        & $AAAA\ BAAB\ ABAA\ BBAA\ ABAA$   & 0.22    \\
5        & $BAAB\ BAAA\ BBBA\ BABA\ ABAB$   & $<0.15$ \\
6        & $AAAB\ BABB\ ABAB\ BABA\ BABA$   & 0.15    \\
\hline
\end{tabular}
\caption{The six sequences studied. The errors in $T_f$ are 
approximately $0.02$.}
\label{tab:1}
\end{center}
\end{table}
For each system ${\cal O}(100)$ Monte Carlo runs, each 
consisting of $3.3\cdot 10^5$ steps were performed. The time averages of 
the mean-square displacement (eq. \ref{d2}) $\la \delta^2 \ra_t$ are 
obtained by averaging over consecutive subsets of conformations with temporal 
extension  $t$ 
(this corresponds to an average time separation of $\approx t/3$).
The simulations were performed at three
different temperatures $T=0.15,0.23$ and $0.4$. 
\begin{figure}[t]
\centering
\mbox{
\psfig{figure=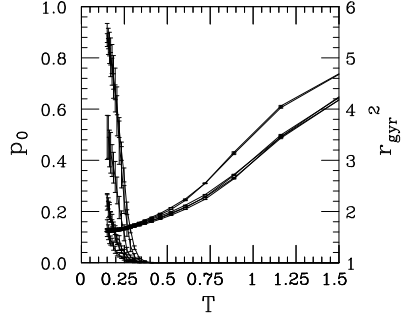,width=5.5cm}
}
\caption{$p_0$ -- the probability for the system to be found close to
the native state, and $r_{gyr}^{2}$ -- the radius of gyration squared, as
a function of temperature.}
\label{rgp0}
\end{figure}
The two lowest correspond to folding temperatures of sequences 3 and 6 
respectively (see Table \ref{tab:1}), whereas  $T=0.4$ is low enough to
correspond to compact states but high enough so that the ground state is almost 
completely de-populated.
This can be seen from Fig.~\ref{rgp0} (data from \cite{Irback:96}),
in which the population of the ground state $p_0$ and the radius of gyration
(squared) is plotted versus temperature.

\begin{figure}[h]
\centering
\mbox{
\psfig{figure=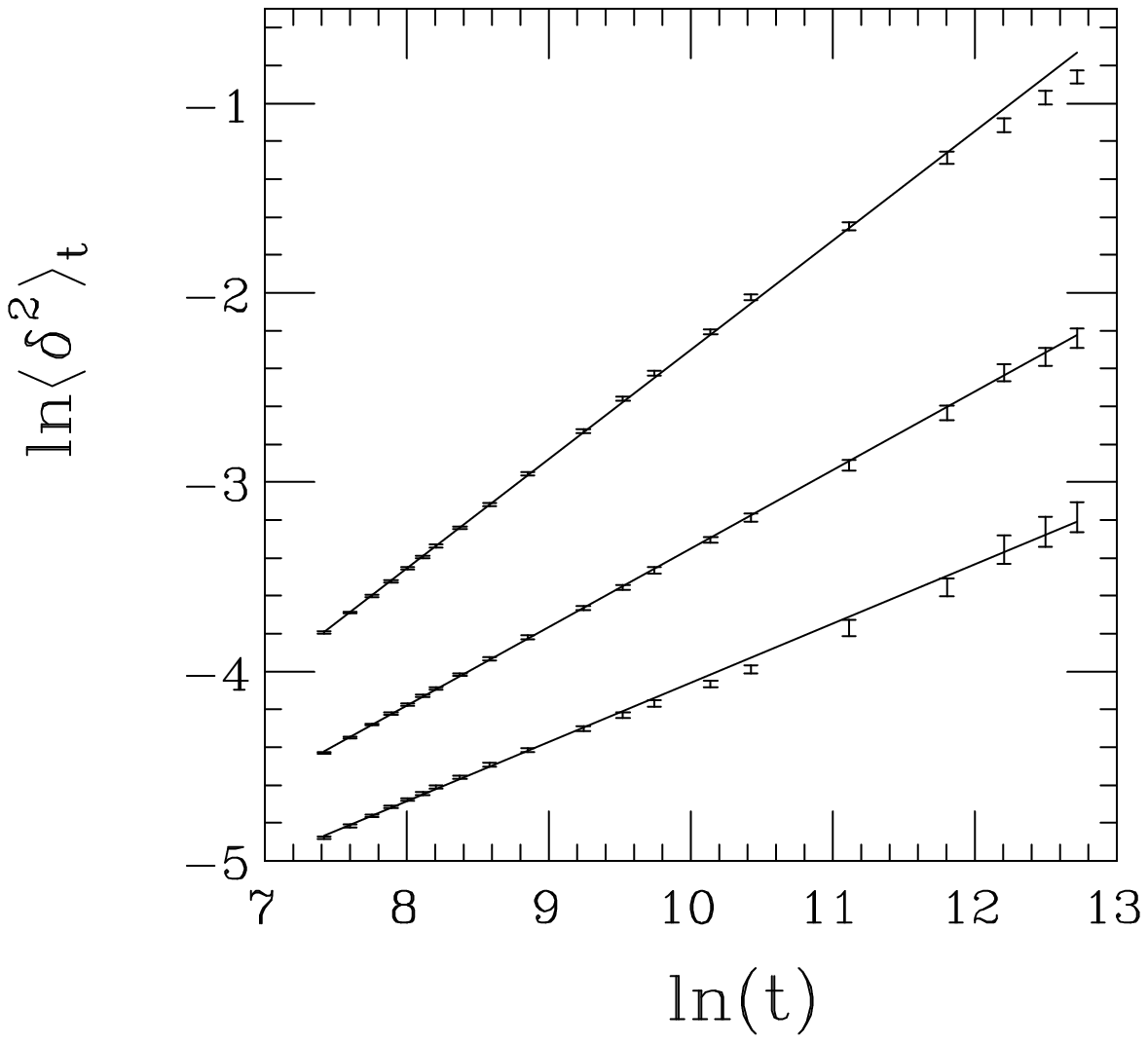,width=4.7cm}
\hspace{1cm}
\psfig{figure=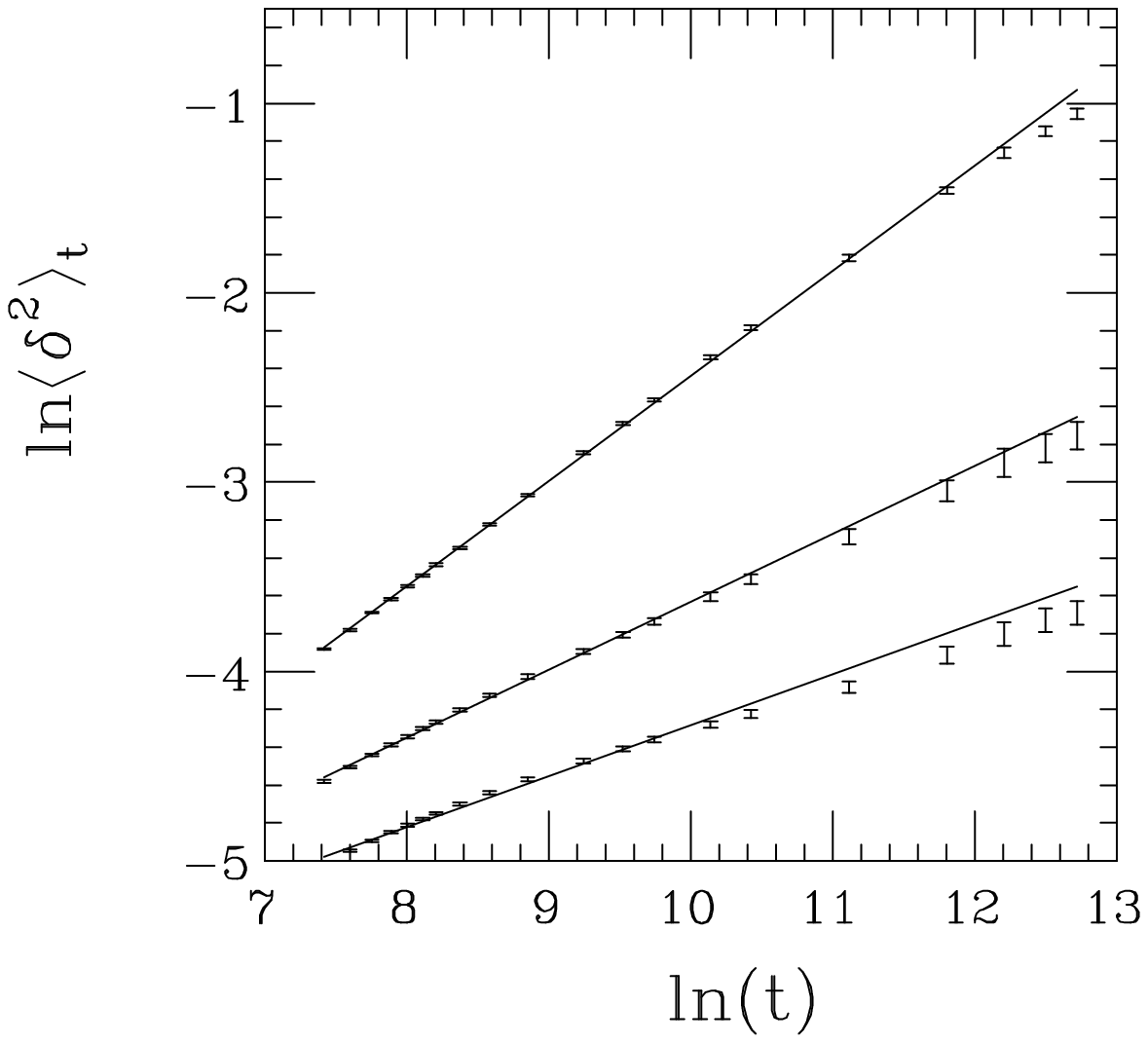,width=4.7cm}
%\hspace{1cm}
%\psfig{figure=\PLOTDIR/d2_t16_v2.ps,width=4.5cm}
}
\caption{$\la\dtwo\ra_t$ as a function of $t$ for sequence no. 1 and 3. The 
data correspond to, from top to bottom, the temperatures $T=0.4, 0.23$ and 
0.15. The solid lines are linear fits.}
\label{d2t}
\end{figure}

In Fig. \ref{d2t}  the time dependence of $\la\delta^2\ra_t$ is displayed for
short times. The data are well parametrized by a 
behavior of the type
\beq
\la\delta^2\ra_t \propto t^{\nu}
\label{nu}
\eeq
with a temperature dependent exponent $\nu$. For a harmonic chain one has 
$\nu=1/2$
while the $T\rightarrow \infty$ limit gives $\nu=1$ \cite{Marinari:92a}.
At short times the high temperature limit is valid also for this model. 
For the sequences here examined $\nu$ grows with temperature (Table \ref{tab:2})
in a roughly sequence independent manner.
At a fixed (absolute) temperature there is a small difference in that the best 
folder is the chain with the slowest kinetics. This can partly be attributed 
to the fact that this system spend a lot of time in the vicinity of the native 
state although this effect mainly should affect the pre factor in eq. \ref{nu}.
On the other hand when comparing the systems at the same ``physical''
temperature the difference is larger and in the other direction. The best folder
now has the fastest kinetics. That is, when comparing the sequences at 
temperatures chosen such that the chains spend an equal amount of time in 
the native state, high thermodynamic stability implies fast kinetics.
%
%a good folder is a chain with fast kinetics. That is, when comparing 
%the sequences at what would correspond to the same physical temperature, 
%high thermodynamic stability implies fast kinetics.
%
%On the other hand when comparing the systems at what would correspond 
%to the same physical temperature i.e. different absolute temperatures, chosen
%such that the chains spend an equal amount of time in the native state, 
%the fact that the kinetic aspects here studied are fairly sequence 
%independent implies that the good folder, i.e. the chain with highest folding 
%temperature, also is the chain with the fastest kinetics.
%
\begin{table}[t]
\begin{center}
\begin{tabular}{|c|c c c|} \hline
no.      &T=0.15 & 0.23 &0.4 \\
\hline	      				   
1        & 0.31 & 0.41 & 0.58 \\
2        & 0.31 & 0.40 & 0.57 \\
3        & 0.27 & 0.36 & 0.56 \\
4        & 0.29 & 0.37 & 0.56 \\
5        & 0.32 & 0.42 & 0.59 \\
6        & 0.30 & 0.40 & 0.58 \\
\hline
\end{tabular}
\caption{The exponent $\nu$ in eq. 5  for the different sequences at 
different temperatures. The errors are approximately $0.03$ }
\label{tab:2}
\end{center}
\end{table}
This means that the crucial requirement to be satisfied for a sequence to 
represent a good folder is thermodynamic stability of the ground state. The more
stable the ground state is the faster becomes the kinetics, at the 
folding temperature, towards this ground state.
This is in line with what was found in \cite{Irback:97}.
With a somewhat different distance measure a similar study was performed 
for the IMP-model in \cite{Marinari:92a}. The investigation was here focused
on the behavior of the exponent $\nu$ at fixed temperature when the strength
$\epsilon$ of the quenched disorder was changed. The results ranged from 
$\nu\approx2/3$ for the ordered system ($\epsilon=0$) to $ \nu \approx1/2$ for 
a chain with strong quenched disorder ($\epsilon=10$). The difference between 
different realizations of the couplings $\eta_{ij}$ at fixed value $\epsilon$
was found to be quite small although no distinction was made between these
realizations in terms of their thermodynamic properties.

\begin{figure}[h]
\centering
\mbox{
\psfig{figure=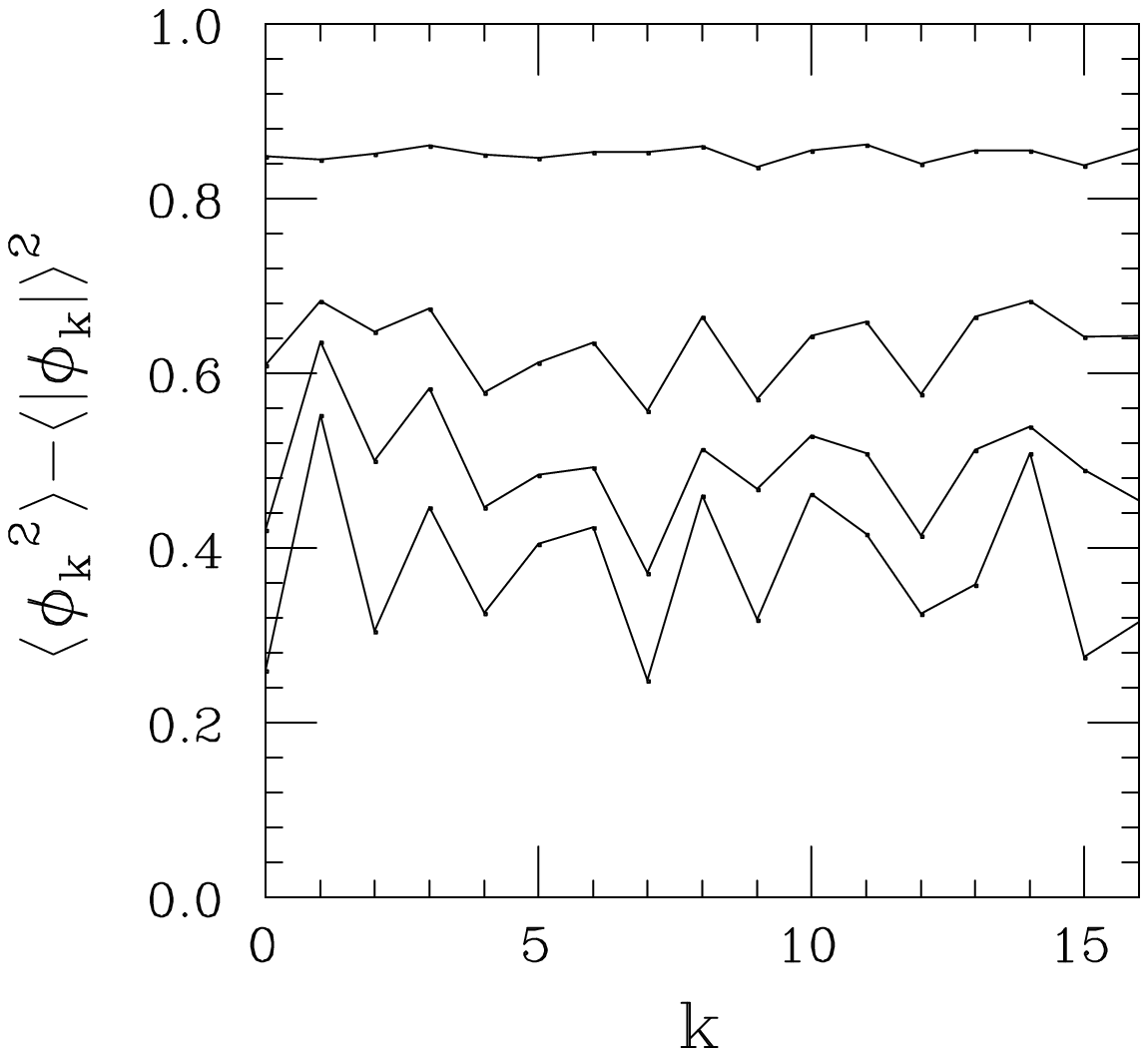,width=4.7cm}
\hspace{1cm}
\psfig{figure=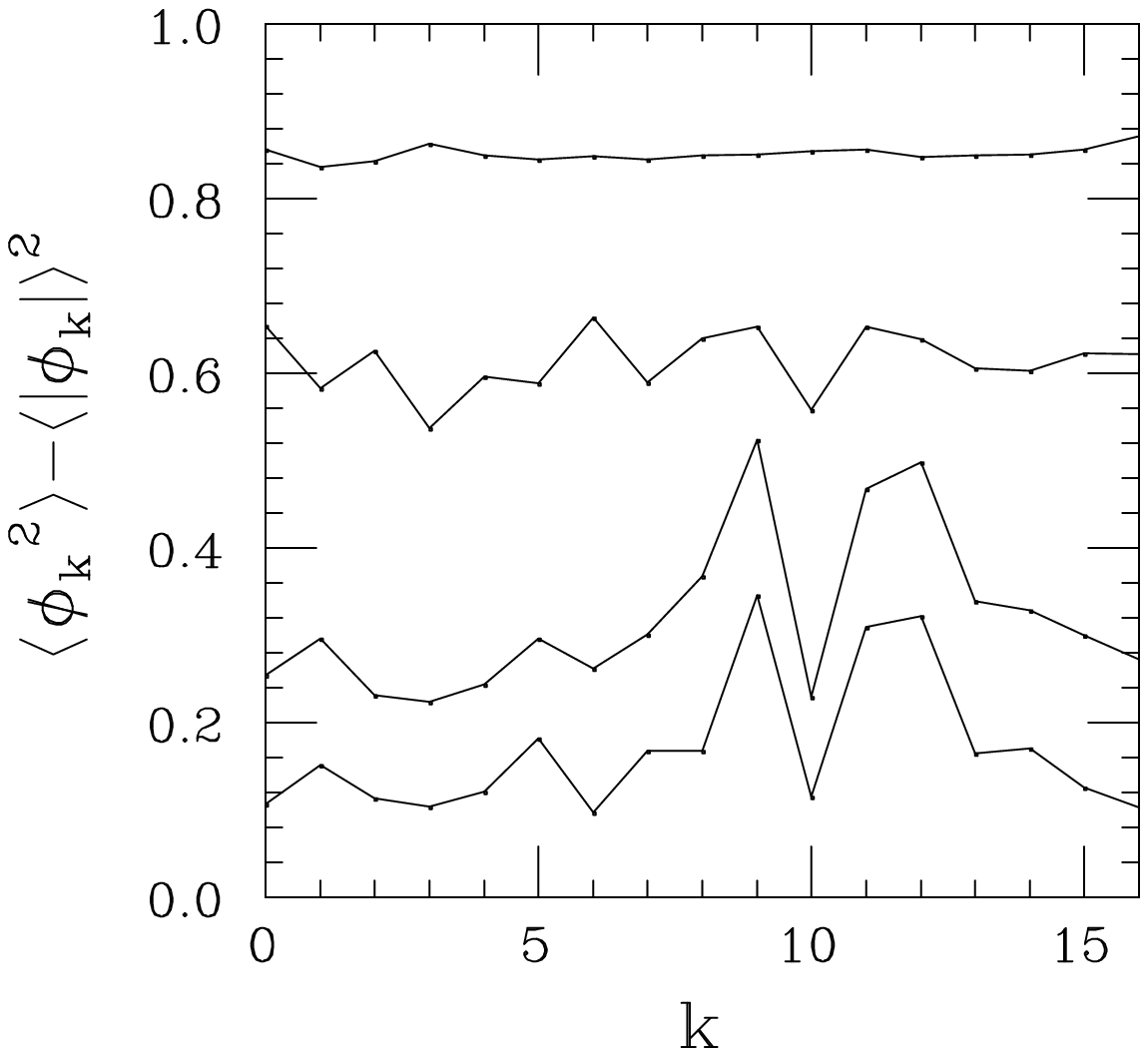,width=4.7cm}
%\hspace{1cm}
%\psfig{figure=\PLOTDIR/tor_16.ps,width=4.5cm}
}
\caption{Fluctuation of torsional angels for sequences no. 1 and 3. The data
corresponds to, from top to bottom, the temperatures 
$T=1.67, 0.41, 0.23$ and  0.15.}
\label{torfu}
\end{figure}

Next we study how the structural fluctuations decay as the temperature is
lowered. All thermal averages where obtained with the ``simulated tempering'' 
method \cite{Marinari:92b}.
The torsional angles are  defined by 
$\cos\phi_k~=~(\bv_k \times \bv_{k+1})\cdot (\bv_{k+1} \times \bv_{k+2})$.
In Fig. \ref{torfu} the fluctuation in these angles, defined by 
$\la \phi^2_k\ra - \la \left| \phi_k \right| \ra^2$ is showed as function of 
temperature. The lines represent a decreasing sequence of temperatures
ranging from $T=1.67$ (top) to $T=0.15$ (bottom). 
As can be seen, the sequence with good folding properties (seq.~3) have a more
drastic ``freeze out'' of the (torsional) degrees of freedom, and at the lowest 
temperature, where $p_0 \approx 0.9$
only a few (3) of the torsion angles have significant fluctuations. For sequence
1 on the other hand  $p_0(T=0.15) \approx 0.25$, \ie the thermodynamic 
stability requirement for the ground state is not yet satisfied, all the angles 
show rather large thermal fluctuations. These measurements do however not 
provide us with any information concerning the amount of correlations present
in this process. In order to examine this issue we estimate an effective
number of degrees of freedom for the chain,
by calculating the eigenvalues to the correlation matrix 
\beq
\rho_{ij} = \la \bv_i \cdot \bv_j \ra %- \la \bv_i\ra\cdot\la \bv_j\ra
\eeq
%
%The trace of the correlation matrix $\rho$ is a measure of the total variance 
%of the system, while the individual eigenvalues represent the variances in 
%the principal directions.
We define an ``effective size'' of the chain as  
\beq
N_{\!\mbox{\em eff}} = \frac{\Tr \rho}{\lambda_0}
\eeq
where $\lambda_0$ is the largest eigenvalue of $\rho$.
\begin{figure}[t]
\centering
\mbox{
\psfig{figure=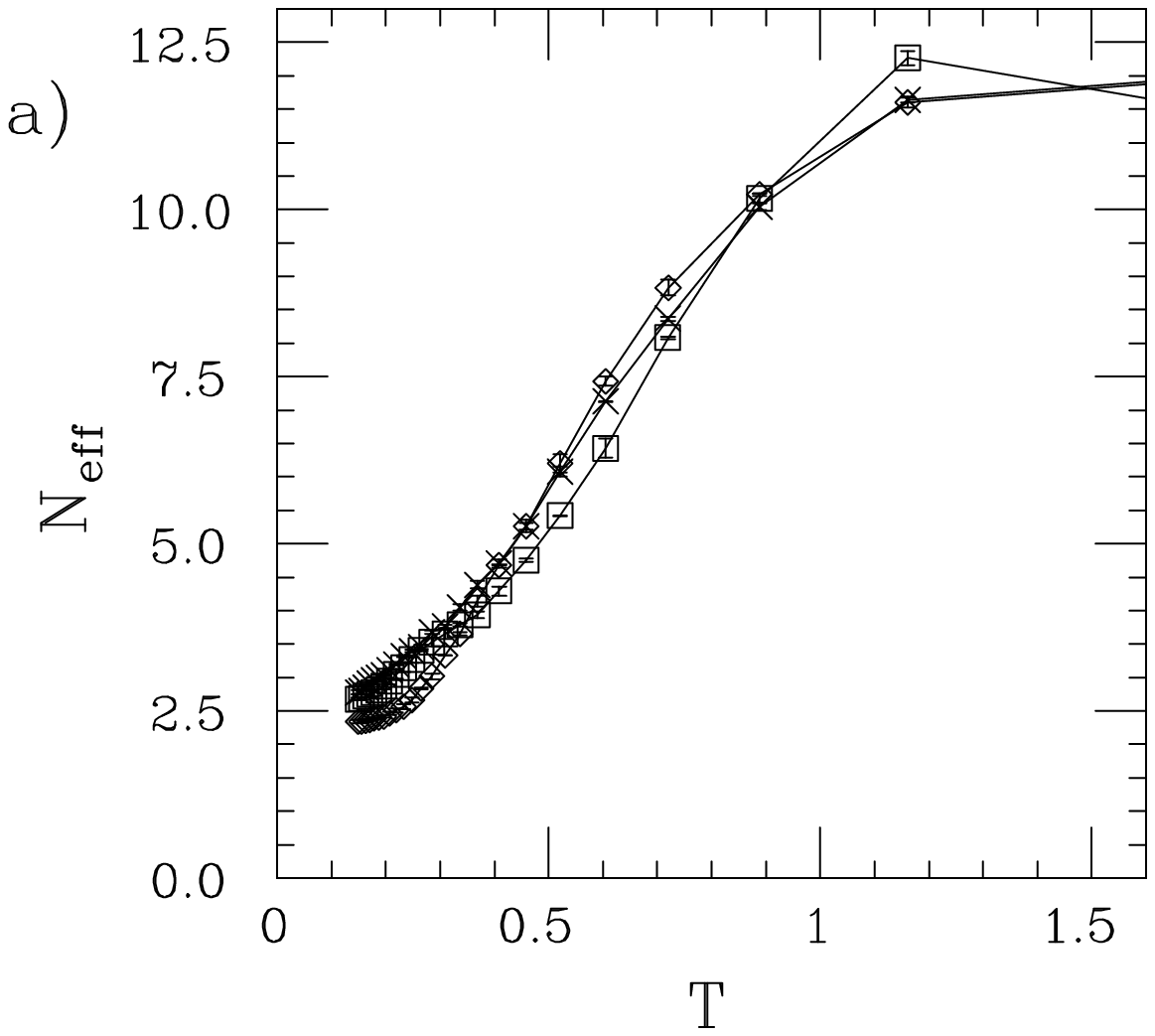,width=4.6cm}
\hspace{1.3cm}
\psfig{figure=sc_final.ps,width=4.6cm}
}
\caption{{\bf a)} The ``effective size'' of the chain
as a function of temperature. 
{\bf b)} The scalar product of the leading eigenvector at $T$ with the one at 
$T=0.15$.
The data correspond to seq. 1 ($\times$),
seq.~7 ($\diamond$) and seq. 16 ($\Box$) respectively. }
\label{neff}
\end{figure}
For this particular choice of correlation matrix the trace equals $N-1$.
As can be seen from Fig.~\ref{neff} a there is a gradual decrease in the 
number of effective degrees of freedom as the temperature is lowered. This is 
in contrast to the behavior expected from a quadratic potential,
for which $N_{\!\mbox{\em eff}}$ would have been temperature independent. 
Thus the energy landscape becomes more and more asymmetric as the temperature
goes down. 
Comparing with the behavior of $r_{gyr}^2$ and $p_0$ in Fig.~\ref{rgp0} we 
see that this ``freezing out'' effect occurs mainly in the compact state but 
before the ground state is populated. 
In order to see how high in temperature the asymmetry defined at the lowest
temperature persists we show in Fig.~\ref{neff} b the scalar product 
between the dominant (normalized) eigenvector at $T=0.15$, $e_0(0.15)$ with the 
corresponding one at temperature $T$,  $e_0(T)$. This represents a measure on 
how similar the direction specifying the largest structural fluctuations
is compared with the principal direction of the ``valley'' hosting the  native
state, or rather the dominating ``valley'' at $T=0.15$. For seq. 3 and 4 at 
least these coincide.
As can be expected at high temperatures there is no memory of this 
direction and although the qualitative behavior is similar for the different 
chains the sequence dependence is rather large compared to that of 
$N_{\!\mbox{\em eff}}$. This reflects the fact that $e_0(0.15)\cdot e_0(T)$ is 
directly related to the properties of the low-end of the energy spectrum 
while $N_{\!\mbox{\em eff}}$ is more general in character.
\section{Conclusions}
We have studied the kinetic behavior of a 3D off-lattice protein 
model in terms of shape diffusion at short time scales in connection with 
folding. We find that sequences with 
very different thermodynamic behavior, e.g. with respect to  the stability of
the ground state, have fairly similar kinetic behavior at short time scales. 
This suggests that the crucial requirements for a chain to have
good folding properties are mainly thermodynamic in nature.
Furthermore, we examine the behavior of the structural fluctuations of the
system --  how large are the correlations in the process of freezing out the  
configurational fluctuations? By estimating an effective number of degrees
of freedom present in the system we find that the collective effects in this 
process are indeed very high and that the asymmetry of the energy landscape is
increased as the temperature is lowered.  
% JOURNALS
\newcommand  {\COSB}    {{\it Curr.\ Opin.\ Struct.\ Biol.\ }}
\newcommand  {\JCP}     {{\it J.\ Chem.\ Phys.\ }}
\newcommand  {\EL}      {{\it Europhys.\ Lett.\ }}
\newcommand  {\JP}      {{\it J.\ Phys.\ A.\ }}
\newcommand  {\PR}      {{\it Phys.\ Rev.\ }}

\end{document}